# Microphase Separation in Random Multiblock Copolymers

E. N. Govorun,[a)] A. V. Chertovich

*Faculty of Physics, Lomonosov Moscow State University, Leninskie Gory 1-2, Moscow 119991 Russia*

Microphase separation in random multiblock copolymers is studied with mean-field theory assuming that long blocks of a copolymer are strongly segregated, whereas short blocks are able to penetrate into "alien" domains and exchange between the domains and interfacial layer. A bidisperse copolymer with blocks of only two sizes (long and short) is considered as a model of multiblock copolymers with high polydispersity in the block size. Short blocks of the copolymer play an important role in microphase separation. First, their penetration into the "alien" domains leads to the formation of joint long blocks in their own domains. Second, short blocks localized at the interface considerably change the interfacial tension. The possibility of penetration of short blocks into the "alien" domains is controlled by the product $\chi N_{sh}$ ($\chi$ is the Flory-Huggins interaction parameter, $N_{sh}$ is the short block length). At not very large $\chi N_{sh}$, the domain size is larger than that for a regular copolymer consisting of the same long blocks as in the considered random copolymer. At a fixed mean block size, the domain size grows with an increase in the block size dispersity, the rate of the growth being dependent of the more detailed parameters of the block size distribution.

## I. INTRODUCTION

Recently, a great progress in methods for the synthesis of random multiblock copolymers has been achieved.[1–8] Such properties of new materials as enhanced mechanical stability,[2–4] sustainability,[3–5] and high proton conductivity[7,8] promise a variety of practical applications. These properties are due to the microphase separation in random multiblock copolymers and, in particular, are related to the appearance of bicontinuous phases.

---

[a)] Author to whom correspondence should be addressed. Electronic mail: govorun@polly.phys.msu.ru.

Random multiblock copolymers are polydisperse in macromolecule length and/or block size. The polydispersity effect on the microphase separation attracts an increased attention last decade.[1–19] New types of polymer components and architectures are tested to produce small-size patterns of different geometries.[20–25] So-called high-$\chi$ materials (with strongly incompatible components) offer the possibility of diblock copolymer sub-10 nm patterning, however, a decrease in the block length definitely means a loss in mechanical properties. Multiblock copolymers could help to overcome this disadvantage, combining small domain sizes with good mechanical properties.

Regular and random linear multiblock copolymers are thoroughly investigated in terms of the weak segregation theory beginning form the classical works[26–30] and different aspects of their polydispersity are studied up to now.[8] In general, a dispersity in composition leads to an increase in the structure period. In the strong segregation theory, microphase separation for regular block copolymers was analyzed using different approaches.[22–25,31–33] Besides, the generalized method of the self-consistent field theory was developed, which permitted calculating phase diagrams for the wider parameter range, describing block conformations in detail and finding a bridge/loop ratio in the strong segregation limit.[34,35] The fraction of "bridges" is not small that is important for understanding melt structure and mechanical properties.[35–39]

The polydispersity effect on the microphase separation was previously investigated for diblock copolymers with the self-consistent field approach and in the strong segregation limit.[2] For multiblock copolymers, this effect was not studied before in the strong segregation theory. Highly polydisperse random multiblock copolymers possess both short and long blocks. The incompatibility $\chi N$-parameter for short blocks could be not large enough for segregation from blocks of another type. The possibility of pulling out the short blocks from their domains is ignored in the self-consistent field approach.

The investigations of random multiblock copolymer melts in computer simulations exhibited rough lamellae or bicontinuous-like structures with the period being very slightly dependent of the incompatibility parameter.[17-18] For random multiblock copolymers obtained via interchain exchange reactions[40] and for the specially created block size distribution (pattern-modified),[19] a considerable amount of "alien" monomer units in lamellae was found in the computer simulations. Besides, an enhanced concentration of short blocks at the interface can be also expected. The experimental data on the content of "alien" components in polystyrene and polybutadiene domains gave the weight fractions from several to more than 10 per cent for alternating and random multiblock copolymers composed of quite long precursors.[2]

In the present paper, we suggest a model of microphase separation in random multiblock copolymers taking into account the ability of short blocks to penetrate into "alien" domains. We consider a special type of random multiblock AB-copolymers with the bimodal block size distribution. That is, the macromolecules consist of alternating sequences of A and B blocks, every block being short or long with a certain probability independently of the other block types. The monomer unit sequences of such copolymers are determined by the long/short block ratio and two block lengths. By varying these parameters, the value of the block size dispersity (or polydispersity index) can be controlled. For the sake of simplicity, a symmetric copolymer composition corresponding to a lamellar melt structure is considered.

Long blocks of A and B types are assumed to be incompatible enough for their strong segregation in different domains, while short blocks can exchange between the interface and "alien" domains. If a short block adjacent to two long blocks penetrates into an "alien" domain, then a joint block composed of two long blocks separated by a short one is confined in that domain with the ends of the joint block localized at the interface. Such joint blocks are not stretched or even unstretched in comparison with usual single long blocks. The elastic free energy of not stretched blocks is calculated analytically as a correction to the free energy of homopolymer chains in a domain due to the chain end localization at the surface. This method is similar to the calculation of the conformational energy of copolymer blocks with both ends localized at a globule surface.[41] To describe a melt structure, we generalize the strong segregation theory approach[27] taking into account the influence of the short block location on the interaction energy, interfacial tension, and conformational free energy.

With our simple model of random multiblock copolymers, we are going to answer the following general questions: i) what factors control the penetration of short blocks into "alien" domains and what could be the value of their volume fraction, ii) how the domain size depends on the block size dispersity and Flory-Huggins parameter.

## II. THE MODEL

We consider a melt of a random multiblock AB copolymer consisting of long and short blocks. A block of any type consists of $N_{sh}$ monomer units with the probability $p_{sh}$ or of $N_{long}$ monomer units ($N_{long}/N_{sh} \gg 1$) with the probability $p_{long}=1-p_{sh}$, the sizes of neighbor blocks being not correlated. For both A and B monomer units, their volumes and sizes along a chain are equal to $\upsilon$ and $a$, respectively. The number average block size is equal to $\overline{N} = p_{long} N_{long} + p_{sh} N_{sh}$ and the dispersity, or polydispersity index (PDI), to $Đ = \dfrac{\overline{N}_w}{\overline{N}} = \dfrac{p_{sh} + p_{long}(N_{long}/N_{sh})^2}{(p_{sh} + p_{long} N_{long}/N_{sh})^2}$. The monomer unit interactions are characterized by the Flory Huggins parameter $\chi$. The total number of blocks in every macromolecule is assumed to be large, and therefore the translation entropy of whole macromolecules and their ends should not be taken into account in the free energy of the system. The individual macromolecules can differ in chain length and block number.

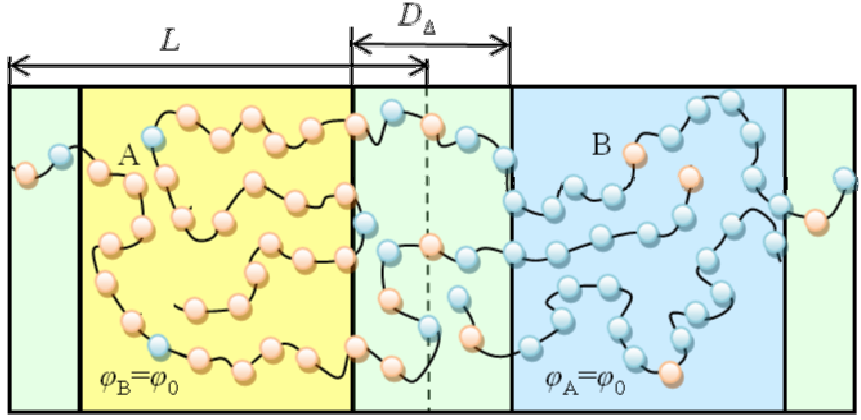

FIG. 1. Schematic of a segregated melt of the random multiblock copolymer with $N_{sh}=1$ and $N_{long}=6$.

If the repulsion between A and B monomer units is strong enough ($\chi N_{long} \gg 1$), then a microphase separation takes place. For the sake of simplicity, it is assumed that a lamellar structure of alternating layers containing mostly A or B monomer units is formed for the symmetric case of 50:50 composition (Figure 1). Long blocks A and B are in the layers of their own type, whereas short blocks can be at the interface between layers or in "alien" domains. It is assumed that only solitary short blocks (adjacent to two long blocks of another type) may penetrate into "alien" domains and short blocks cannot stay in the layers of their own type because in that case two adjacent blocks would be in the "alien" layers. That is, a short block with at least one adjacent short block should always be at the interface.

The volume fraction of short A blocks in B layers, $\varphi_A$, and of short B blocks in A layers, $\varphi_B$, are equal to each other, $\varphi_A=\varphi_B=\varphi_0$ ($\varphi_0 \ll 1$). The layers A and B have the same thickness, $L$, and the interfacial layer thickness is denoted by $D_\Delta$. The average volume fraction of monomer units of short blocks of any type is equal to $\varphi_{sh} = \dfrac{p_{sh} N_{sh}}{p_{sh} N_{sh} + p_{long} N_{long}}$. Then, the fraction of short blocks in "alien" layers is equal to $p_0 = \varphi_0/\varphi_{sh}$. Let $\Delta$ be the thickness of the layer consisting of only short blocks located at the interface. From the normalization condition with respect to the number of short blocks, $\varphi_{sh}L=\varphi_0 L+\Delta(1-\varphi_0)$, so that

$$\Delta = L\frac{\varphi_{sh} - \varphi_0}{1 - \varphi_0}. \tag{1}$$

The volume fraction of short blocks in "alien" domains, $\varphi_0$, can change from 0 to the maximum value, $\varphi_{0max}$, for which all solitary short blocks (adjacent to two long blocks) are in "alien" domains. The maximum number fraction of short blocks in "alien" domains is equal to the fraction of solitary blocks, $p_{0max} = p_{long}^2$. The maximum volume fraction is equal to $\varphi_{0max} = p_{long}^2 \varphi_{sh}$ and correspondingly the thickness $\Delta$ takes the minimal value $\Delta_{min} = L\frac{(1 - p_{long}^2)\varphi_{sh}}{1 - p_{long}^2 \varphi_{sh}}$. The dependence of the maximum volume fraction of short blocks in "alien" domains, $\varphi_{0max}$, on the number fraction of short blocks is presented in Figure 2. The maximum value of $\varphi_{0max}$ corresponds to the number fraction of short blocks $p_{sh} \approx 0.45$ at $N_{long}/N_{sh}=5$ and tends to 0.5 if $N_{long}/N_{sh} \to \infty$. At $N_{long}/N_{sh} \gg 1$, it does not exceed several per cent.

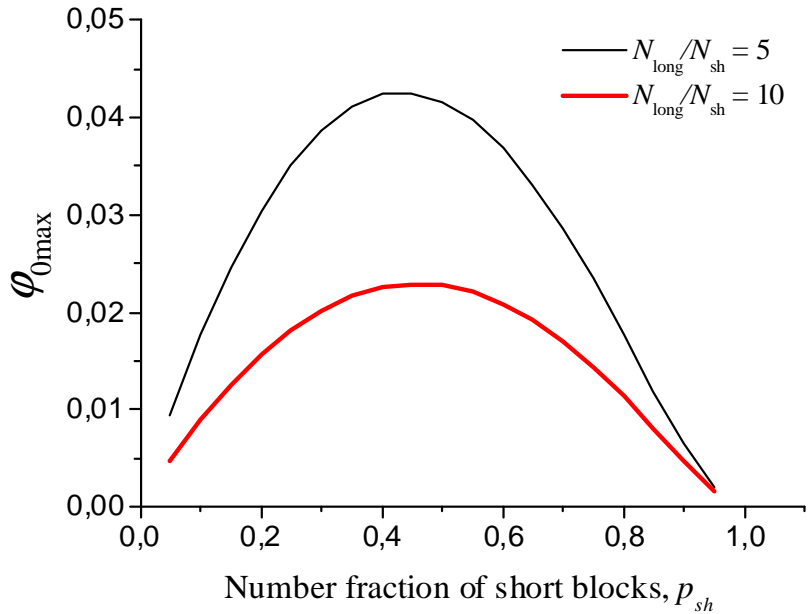

FIG. 2. Maximum volume fraction of short blocks in "alien" domains, $\varphi_{0max} = p_{long}^2 \varphi_{sh}$, vs the number fraction of short blocks, $p_{sh}$, at $N_{long}/N_{sh}=5$ (thin black curve) and $N_{long}/N_{sh}=10$ (thick red curve). Only solitary short blocks (adjacent to two long blocks) may penetrate into "alien" domains.

The presence of short blocks at the interface between A and B layers should enlarge the interfacial layer thickness in comparison with that in a melt of the multiblock copolymer with long blocks only. We assume that the interfacial thickness, $D_\Delta$, is equal to the sum of the interfacial thickness in two limiting cases: $D_\Delta = \Delta + D$, where $\Delta$ is given by Eq. (1) and $D$ is the interfacial thickness for a model solution of the multiblock copolymer consisting of only long blocks in a nonselective solvent at the constant solvent volume fraction equal to $\varphi_0$. For this solution, the Flory-Huggins parameter describing interactions between the solvent and monomer units of both types is taken to be equal to $\chi$. Interactions of the polymer with such solvent mimic the volume interactions between long and "alien" short blocks in the considered random multiblock copolymer melt. The characteristic thickness of the interface between A and B domains in such solution is equal to (see **Appendix A**)

$$D = D_0 (1-\varphi_0)^{1/2} / (1-2\varphi_0), \qquad (2)$$

where $D_0 = a/\sqrt{6\chi}$ is the interface thickness in an A and B homopolymer melt.

It can be expected that the value of the interfacial thickness $D_\Delta$ is determined mainly by the value of $D$ at weak incompatibility of A and B blocks, $\Delta \ll D$, since the thickness $D_0$ is quite large in that case and some part of short blocks are in "alien" domains, which have not very large size $L$. For strongly incompatible blocks (high $\chi$), the interfacial thickness $D_\Delta$ is determined by the thickness $\Delta$ (1), $\Delta \gg D$, since the thickness $D_0$ is small and short blocks are pushed out of "alien" domains ($\varphi_0 \approx 0$) of quite large thickness $L$.

### A. Free energy of a homogeneous melt

To control whether the microphase separation takes place in a random multiblock copolymer melt, the free energies of the homogeneous melt and the melt with a lamellar structure should be calculated and compared with each other. The volume fraction of A and B monomer units in such melt is equal to $\bar\varphi_A = \bar\varphi_B = 1/2$. It is assumed that short blocks are mobile and several neighboring short blocks can move together. The free energy $F_0$ is equal to a sum of the interaction energy and translational entropy contribution of short blocks (in the form of an ideal-gas free energy),

$$\frac{F_0 v}{k_B TV} = \chi \bar{\varphi}_A \bar{\varphi}_B + \frac{\varphi_{sh}}{N_{sh}} p_{long}^2 \sum_{i=1}^{\infty} p_{sh}^{i-1} \ln \frac{\varphi_i}{e}. \qquad (3)$$

Here $V$ is the volume of the system, $T$ is the thermodynamic temperature, $k_B$ is the Boltzmann constant, $\varphi_i$ is the volume fraction of monomer units belonging to sequences containing $i$ short blocks, which are adjacent to two long blocks at the edges. The number of solitary short blocks with two adjacent long blocks is equal to $M p_{sh} p_{long}^2$, where $M$ is the total number of blocks. The number of pairs of short blocks is equal to $M p_{sh}^2 p_{long}^2$ and so on. Then, $\varphi_i = i p_{sh}^{i-1} p_{long}^2 \varphi_{sh}$, $\varphi_{sh} = \sum_{i=1}^{\infty} \varphi_i$.

**B. Free energy of a layered melt**

We assume that long polymer blocks are strongly segregated ($D_\Delta \ll L$), whereas short blocks are in equilibrium between the layers and interfacial region. The free energy of the system, $F$, depends on the volume fraction of short blocks in the layers, $\varphi_0$. It includes several contributions,

$$F = E + F_{el} + F_s, \qquad (4)$$

where $E$ is a sum of the interaction energy and translational entropy contribution of short blocks in "alien" layers, $F_{el}$ is the elastic free energy of long blocks, and $F_s$ is the interface free energy. The first contribution has the form

$$\frac{Ev}{k_B TV} = \chi \varphi_0 (1-\varphi_0) + \frac{\varphi_0}{N_{sh}} \ln \frac{\varphi_0}{e}. \qquad (5)$$

For calculating the elastic free energy of long blocks, it is necessary to take into account that the conformational constraints for long blocks separated by $i$ solitary short blocks in an "alien" layer are the same as for a joint block of length $N_i = N_{long} + i(N_{long} + N_{sh})$. This block is localized in a layer of thickness $L$ with the ends located at its surface. Let $p_b = p_{sh} p_0$ be the probability to find a short block located in an "alien" layer. Then, the number of long blocks of length $N_0 = N_{long}$ with both ends at the interface is equal to $Mp_{long}(1-p_b)^2$, the number of joint blocks of length $N_1$ is equal to $Mp_{long}^2 p_b(1-p_b)^2$, and so on. The elastic free energy of blocks, $F_{el}$, can be written as

$$\frac{F_{el}v}{k_B TV} = \frac{1}{\bar{N}} p_{long}(1-p_b)^2 \sum_{i=0}^{\infty} (p_{long} p_b)^i f_{bl}(N_i, L), \tag{6}$$

where $f_{bl}(N,L)$ is the elastic free energy of a block consisting of $N$ monomer units, which is located inside of the layer of thickness $L$ with the both ends at its surface.

If the thickness $L$ is less or comparable to the characteristic size $L_{0i} = \sqrt{N_i} a$ of a block of length $N_i$, this block is not stretched and the conformational restrictions are caused only by the block end location at the interface. For the calculation of the elastic free energy of joint blocks in that case, we consider it as a correction to the free energy of such blocks in a homogeneous melt as was done earlier for the calculation of the conformational free energy of a polymer block in a globule with both ends at the surface.[41] The elastic free energy of a block calculated using this approach is denoted by $f_L$ (see **Appendix B**). The possibility for a block to take a "loop" (with both ends at the same surface of the layer) or "bridge" (with the ends at the different surfaces of the layer) conformation is taken into account. For quite high block stretching, we assume that the block elastic free energy, $f_{bl}$, tends to the elastic free energy $f_{gs} = \frac{3k_B T}{2} \frac{L^2}{N_i a^2}$ of Gaussian chains. At the intermediate values of $L$ in the range $L_0 < L < L_{sm}$, we use a smoothing approximation, $f_{sm}$, for the dependence of the block elastic free energy, $f_{bl}$, on $L$, which provides continuous dependences of the function, $f_{bl}$, and its derivatives, $\partial f_{bl}/\partial L$, $\partial^2 f_{bl}/\partial L^2$ on $L$:

$$f_{bl}(N_i, L) = \begin{cases} f_L, & L \leq L_0 \\ f_{sm}, & L_0 < L < L_{sm} \\ f_{gs}'' \cdot (L - L_{str})^2/2 + c_{str}, & L > L_{sm} \end{cases} \tag{7}$$

where $L_0 = \sqrt{N_i}a$, the value of $L_{sm}$ is taken to be slightly larger than $L_0$,

$$f_{sm}(N_i, L) = f_0 + f_0'(L - L_0) + \frac{1}{2}f_0''(L - L_0)^2 + \frac{f_{gs}'' - f_0''}{6(L_{sm} - L_0)}(L - L_0)^3, \; f_0 = f_L(N_i, L_0), \; f_0' = \left.\frac{\partial f_L}{\partial L}\right|_{L=L_0},$$

$$f_0'' = \left.\frac{\partial^2 f_L}{\partial L^2}\right|_{L=L_0}, \; f_{gs}'' = \frac{\partial^2 f_{gs}}{\partial L^2} = \frac{3k_BT}{N_i a^2}, \; L_{str} = L_{sm} - \frac{f_0'}{f_{gs}''} - \frac{1}{2}\left(1 + \frac{f_0''}{f_{gs}''}\right)\left(\frac{L_{sm}}{L_0} - 1\right), \; c_{str} = f_{sm}(N_i, L_{sm})$$

$$-\frac{f_{gs}''}{2}(L_{sm} - L_{str})^2.$$

The interfacial energy can be represented as the sum[33]

$$F_s = (\sigma + \sigma_e)S, \tag{8}$$

where $\sigma$ is the interfacial tension arising from the elastic energy of long blocks and interaction energy of monomer units in the interfacial region and $\sigma_e$ is the contribution of block end localization at the interface. These terms are taken in the form (see **Appendix A**)

$$\tilde{\sigma} = \frac{\sigma v}{k_B T a} = \frac{a}{12}\frac{1 - \varphi_0}{D_\Delta} + \frac{\chi}{2}(1 - 2\varphi_0)^2 \frac{D_\Delta}{a}, \tag{9}$$

and

$$\sigma_e = \sigma_{e,long} + \sigma_{e,sh}, \tag{10}$$

$$\frac{\sigma_{e,long}}{k_B T} = \frac{M\tilde{p}_{long}}{S}\ln\left(\frac{\tilde{p}_{long}}{\overline{N}e}\frac{L}{\pi D_\Delta}\right), \quad \tilde{p}_{long} = \frac{(1 - p_b)p_{long}^2}{1 - p_{long}p_b}$$

$$\frac{\sigma_{e,sh} S v}{k_B T V} = \frac{\varphi_1 - \varphi_0}{N_{sh}}\ln\left(\frac{\varphi_1 - \varphi_0}{e}\frac{L}{\pi D_\Delta}\right) + \frac{\varphi_{sh}}{N_{sh}}p_{long}^2\sum_{i=2}^{\infty}p_{sh}^{i-1}\ln\left(\frac{\varphi_i}{e}\frac{L}{\pi D_\Delta}\right),$$

$\sigma_{e,long}$ and $\sigma_{e,sh}$ are the contributions of one-end localization for joint blocks with a long next block and of short block localization, respectively. It is taken into account that $V=SL$. The first term in the expression for $\sigma_{e,sh}$ describes solitary short blocks in the interfacial layer.

The equilibrium parameters of the system correspond to the free energy minimum with respect to the volume fraction of short blocks in the layers, $\varphi_0$, and the layer thickness, $L$.

### C. Free energy of a regular copolymer layered melt

To reveal the effect of short blocks, it is illustrative to find the layer thickness and compare the free energy contributions for the melts of a random multiblock copolymer with long and short blocks and of a regular copolymer with long blocks of length $N$ only. The free energy of a regular copolymer melt, $F_{reg}$, consists of the elastic free energy of blocks, $F_{el(reg)}$, and the interfacial free energy, $F_{s0}$,

$$F_r = F_{el(reg)} + F_{s0} \tag{11}$$

According to the calculation of the elastic free energy for a random multiblock copolymer, the elastic free energy for a regular copolymer can be written as

$$\frac{F_{el(reg)} \upsilon}{k_B T V} = \frac{1}{N} f_{bl}(N, L_r).$$

The interfacial free energy can be written as[33]

$$F_{s0} = (\sigma_0 + \sigma_{e0})S, \quad \frac{\sigma_0}{k_B T} = \frac{a}{\upsilon}\sqrt{\frac{\chi}{6}}, \quad \frac{\sigma_{e0}}{k_B T} = \frac{M_r}{S} \ln\left(\frac{M_r \upsilon}{S e \pi D_0}\right),$$

where $\sigma_0$ is the interfacial tension in a melt of A and B homopolymers, $\sigma_{e0}$ is the contribution of localizing block junction points in the interfacial region of thickness $D_0 = a/\sqrt{6\chi}$, $M_r$ is the number of block junction points which is approximately equal to a total number of blocks, $M$, for long enough multiblock chains. Minimizing the free energy, $F_{reg}$ (Eq. (11)), with respect to the layer thickness, $L_r$, and taking into account that $V = MN\upsilon = L_r S$ the equilibrium value of the layer thickness can be found.

## III. RESULTS AND DISCUSSIONS

Penetration of short blocks into "alien" layers can be energetically favorable for two reasons, an increase in the translational entropy of short blocks and appearance of joint long blocks. The dependence of the elastic free energy of joint blocks on the layer thickness calculated according to Eq. (7) and **Appendix B** is presented in Figure 3. For the value $L_{sm} = 1.2 L_0$ taken in the calculations, the parameter $L_{str}/L_0$ was equal to 0.825. The elastic free energy grows slowly (the increase is less or comparable with $k_BT$) with the layer thickening from several monomer unit sizes up to $L \approx L_{0i} = \sqrt{N_i}a$. At $L > L_{0i}$, the elastic free energy increases more rapidly and the free energy gain corresponding to the appearance of a joint block of size $N_1$ instead of two blocks of size $N_0 = N_{long}$ can exceed several $k_BT$.

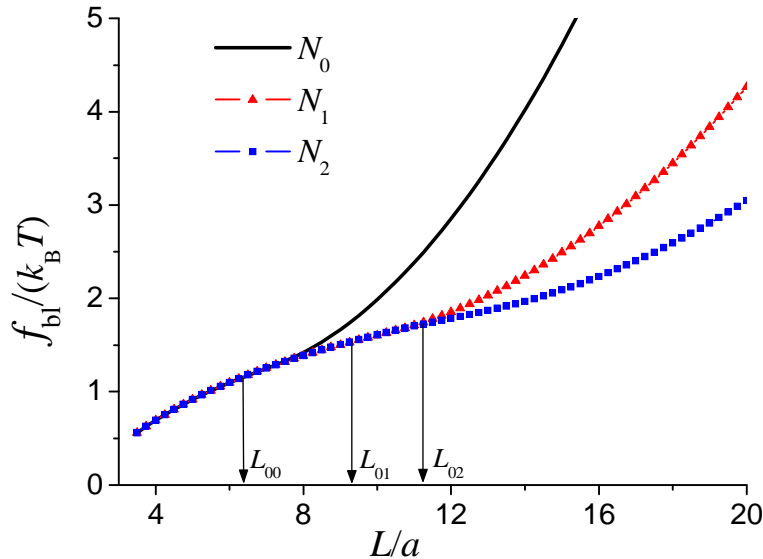

FIG. 3. The block elastic free energy, $f_{bl}$, in units $k_BT$, vs the layer thickness, $L$, for joint blocks of size $N_i = N_{long} + i(N_{long} + N_{sh})$ at $N_{long} = 40$, $N_{sh} = 4$ ($L_{0i} = \sqrt{N_i}a$, $L_{00} = \sqrt{N_{long}}a \approx 6.3a$, $L_{01} \approx 9.2a$, $L_{02} \approx 11.3a$).

The "bridge" fraction for not stretched blocks of a fixed length $k$ is proportional to the number of possible block conformations in a layer and can be estimated as $G_2/(G_1+G_2)$, where $G_1(k,L)$ and $G_2(k,L)$ are the statistical weights of loops and bridges, respectively (see Appendix B). The bridge fraction decreases with $L$ from 0.5 at $L \sim a$ to approximately 0.3 at $L = L_0$. The present model does not permit to analyse a bridge fraction for stretched blocks. However, the fraction of bridges for a random copolymer with a given mean block length will be much less than that for copolymers with monodisperse blocks of the same length. It is because short blocks mostly form loops whereas long blocks occupy the middle part of domains and may form bridges of loops. This picture is consistent with a very low bridge fraction (several per cent) for pattern-modified random multiblock copolymers obtained via layer marking in a homogeneous melt.[19]

In theory, a loop is considered as two linear chains and equal probabilities of bridge and loop conformations are usually assumed.[22–25,31] The more detailed self-consistent field calculations predict the bridge fraction 0.4 for stretched monodisperse middle blocks.[34,35] Note that the combinatorial distributions of bridges and loops in the layers are impotrant for the final structure of multiblock copolymers with not large numbers of blocks in macromolecules, as shown for the copolymers with monodisperse blocks forming usual lamellar structure[31] and for the copolymers with long end blocks and many more short middle blocks forming lamellar-in-lamellar structures.[22–25]

The change in the elastic behavior of the copolymer blocks at $L=L_{00}$ leads to the different character of the dependences of the volume fraction of short blocks in the "alien" domains and of the layer thickness on the block size and Flory-Huggins parameter for not stretched ($L<L_{00}$) and stretched blocks ($L>L_{00}$), as shown in Figures 4 and 5. All equilibrium parametes are calculated by minimizing the free energy $F$ (Eqs. (4)-(10)) with respect to $\varphi_0$ and $L$ and the free energy $F_r$ (Eq. (11)) with respect to $L_r$. Note that the free energy of a homogeneous melt calculated in accordance with Eq. (3) is larger than the free energy of a melt with lamellar structure for all considered values of the parameters.

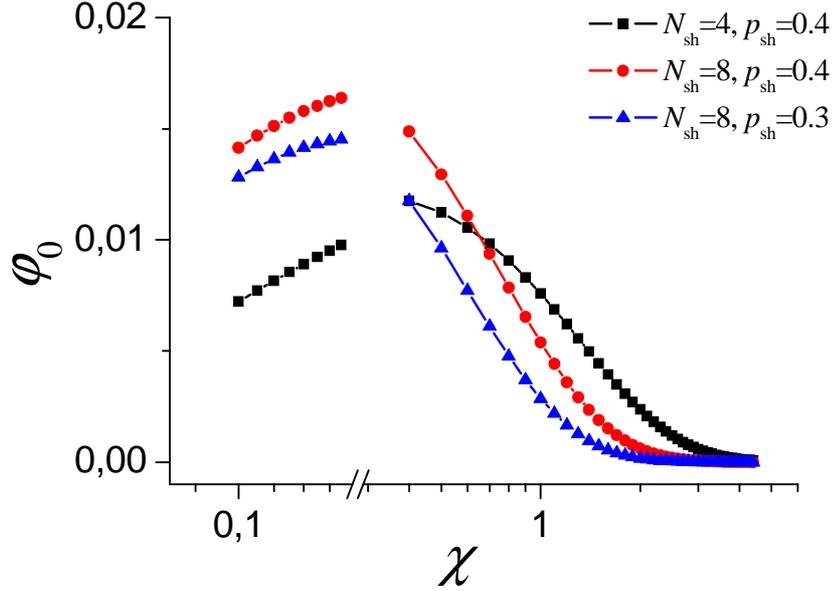

FIG. 4. Volume fraction of the solitary short blocks in "alien" layers, $\varphi_0$, vs the Flory-Huggins parameter, $\chi$, for the random multiblock copolymer with $N_{sh}=4$, $p_{sh}=0.4$ (black curve with squares), $N_{sh}=8$, $p_{sh}=0.4$ (red curve with circles), $N_{sh}=8$, $p_{sh}=0.3$ (blue curve with triangles); the long block size $N_{long}=40$.

The volume fraction of short blocks in "alien" domains, $\varphi_0$, at small $\chi$ (at $L<L_{00}$) is around 0.8% for $N_{sh}=4$ and 1.5% for $N_{sh}=8$ for the multiblock copolymers considered in Figure 4. Approximately one third of solitary short blocks penetrated in "alien" domains, and a larger value of volume fraction $\varphi_0$ corresponds to a larger value of the maximum volume fraction $\varphi_{0max}$ (Fig. 2). Surprisingly, the volume fraction of short blocks slightly increases with $\chi$ that can be related to the rapid growth of the layer thickness, whereas the increase of the layer thickness at a fixed $\chi$ should lead to the penetration of more short blocks in the "alien" layers. For larger $\chi$ ($L>L_{00}$), the monomer unit interactions become to play a more important role in comparison with the entropic factors and the increase in $\chi$ leads to pushing "alien" short blocks out of the layers toward the interface, so that $\varphi_0$ decreases. Longer short blocks are pushed out of "alien" domains at lower values of the Flory-Huggins parameter $\chi$, the maximum amount of short blocks in "alien" domains is observed at $L \approx L_{00}$ and it tends to zero at $\chi N_{sh}>10$.

The weight fraction of "alien" monomer units in the polystyrene and polybutadiene domains was measured for regular and random multiblock copolymers composed of quite long precursors.[2] This fraction varies from several to more than 10 per cent and it is larger for alternating multiblock copolymers than for the random ones, which can be explained from the present work standpoint by a smaller weight fraction of the shortest (solitary) blocks in the random multiblock copolymers. In computer simulations of the microphase separation in a melt of random multiblock copolymers of the special type (pattern-modified),[19] the observed volume fraction of "alien" blocks was equal to 5-7%. The lower volume fraction of "alien" blocks in the present study can be related to the smaller volume fraction of short (solitary) blocks in the considered bidisperse multiblock copolymers.

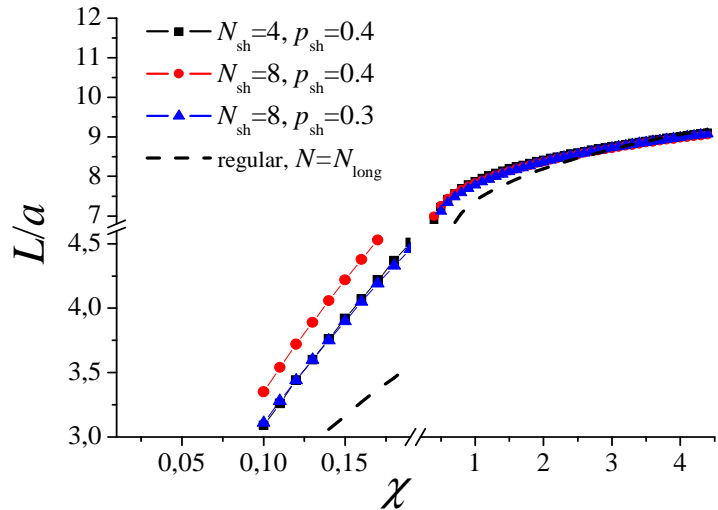

FIG. 5. Layer thickness, $L$, vs the Flory-Huggins parameter, $\chi$, for the random multiblock copolymers with $N_{sh}=4$, $p_{sh}=0.4$ (black curve with squares), $N_{sh}=8$, $p_{sh}=0.4$ (red curve with circles), and $N_{sh}=8$, $p_{sh}=0.3$ (blue curve with triangles); the long block size, $N_{long}=40$. The dashed curve describes the layer thickness for the regular block copolymer with the block size $N_{long}=40$.

Much lower elasticity of unstretched long blocks corresponds to a more rapid growth of the layer thickness, $L$, with the Flory-Huggins parameter, $\chi$, at $L < L_{00}$ than at $L > L_{00}$ (Figure 5). If a certain amount of short blocks is in the "alien" layers, then the layer thickness, $L$ (as well as the interface thickness, $D_\Delta$), increases with the growth of the short block size, $N_{sh}$, at the fixed number fraction of short blocks, $p_{sh}$, (black and red curves) and with the growth of this fraction at the fixed short block size (blue and red curves). Thus, the layer thickness increases with the short block volume fraction, $\varphi_{sh}$, at fixed $\chi$ and $N_{long}$.

The layer thickness for the random multiblock copolymers markedly exceeds the layer thickness for a regular block copolymer with the block size $N=N_{long}$ (dashed curve). This effect can be explained by the penetration of short blocks into the "alien" layers (the size of joint long blocks effectively increases) and by the presence of other short blocks at the interface. For regular multiblock copolymers, the scaling dependence of the layer thickness on $\chi$ and $N$ can be estimated from the analysis of Eq. (11) for the free energy. The elastic block free energy at $L_r \ll \sqrt{N}a$ is approximately equal to $f_{bl}(N,L_r) \approx -k_B T \ln((G_1+G_2)a) \approx k_B T \ln(2L_r/a)$, the contribution of a block end localization at the interface for a regular block copolymer is approximately equal to $k_B T \ln(L_r/D_0)$, and the main interfacial contribution is equal to $\sigma_0 V/L_r$. Minimizing the corresponding free energy expression $k_B T M(\ln(2L_r/a)+\ln(L_r/D_0))+\sigma_0 V/L_r$ with respect to $L_r$ gives the estimate of the equilibrium value

$$L_r \approx 0.5\sigma_0 N \upsilon = \sqrt{\chi/24}\, Na, \qquad (12)$$

which increases linearly with the block length and as $\chi^{1/2}$ with the Flory-Huggins parameter.

For the random multiblock copolymers, the number of joint blocks is 10-20% less than the total number of long blocks if a considerable amount of solitary short block is located in the "alien" layers, since the probability that a long block is the part of a joint block is approximately equal to $p_b p_{long} \sim p_{sh}(1-p_{sh})$. The low elasticity of joint blocks diminishes the overall elastic response and promotes the layer thickening. At the same time, the short blocks localized at the interface can considerably enlarge the interface thickness, $D_\Delta$, and the interfacial tension, $\sigma$ (the second term in Eq. (9) is proportional to $D_\Delta$). Both factors promote an essential increase in the equilibrium layer thickness in comparison with that of regular block copolymers. The critical value of $(\chi N)_{cr} \approx 24$ separating the regimes of stretched and unstretched blocks for a regular block copolymer can be estimated from Eq. (12) and $L_r = \sqrt{N}a$. For a random block copolymer, the critical value should be several times lower.

The analysis of the calculated data leads to the conclusion that the layer thickness at $L>L_{00}$ is mainly controlled by the interaction parameter, $\chi$, the long block size, $N_{long}$, and the mean block size, $\overline{N}$, which is inversely proportional to the total number of polymer blocks. The dependences of the layer thickness on $N_{long}$ for the fixed mean block size but different short block sizes are very close (open and solid symbols in Figure 6). The maximum volume fractions of short blocks, $\varphi_{0max}$, and the block size dispersity (or PDI), $Đ$, for these series are different, nevertheless the layer thickness is almost the same. Note that we consider the fraction of short blocks less than 0.7, otherwise, the total volume fraction of short blocks becomes not small and the condition $D_\Delta << \sqrt{N_{long}}\,a$ for the strong segregation regime can be disturbed.

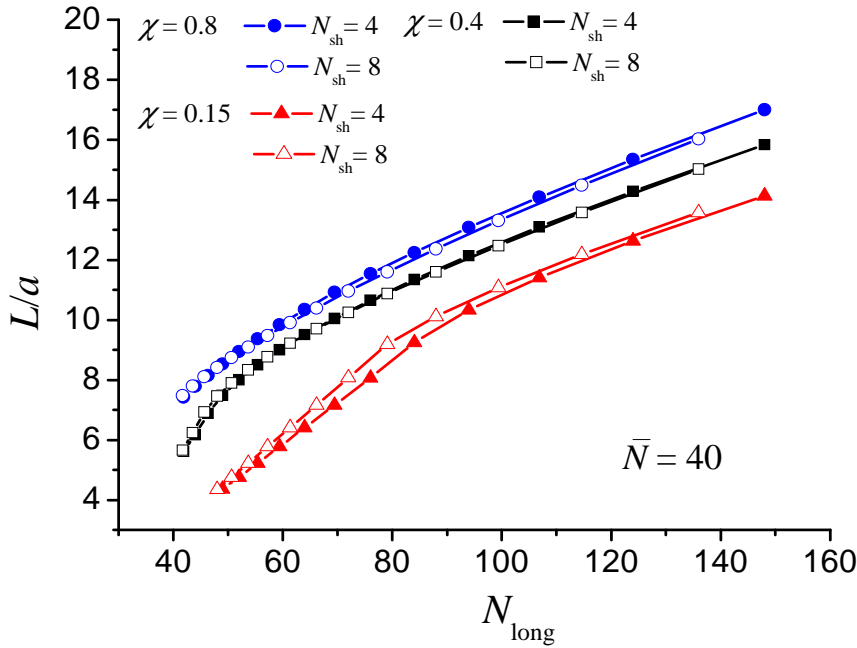

FIG. 6. Layer thickness, $L$, vs the long block size, $N_{long}$, at the fixed mean block size, $\overline{N}=40$ for $N_{sh}=4$ (solid symbols) and $N_{sh}=8$ (open symbols) and for $\chi=0.15$ (triangles), 0.4 (squares), and 0.8 (circles). The range of $N_{long}$ values corresponds to changing the number fraction of short blocks, $p_{sh}$, from 0.05 to 0.65.

The dependences of the layer thickness for the random multiblock copolymers on the dispersity in the block size, $Đ=(p_{sh}N_{sh}^2 + p_{long}N_{long}^2)/\overline{N}^2$, at the same mean block size, $\overline{N} = p_{long}N_{long} + p_{sh}N_{sh} = 40$, and at the fixed size, number fraction, or volume fraction of short blocks are presented in Figure 7. The layer thickness considerably increases with the dispersity at fixed $N_{sh}$ or $\varphi_{sh}$ that stems from an increase in the long block size, $N_{long}$. At fixed $p_{sh}$, the long block length increases only slightly with the dispersity that corresponds to a slight increase in the layer thickness.

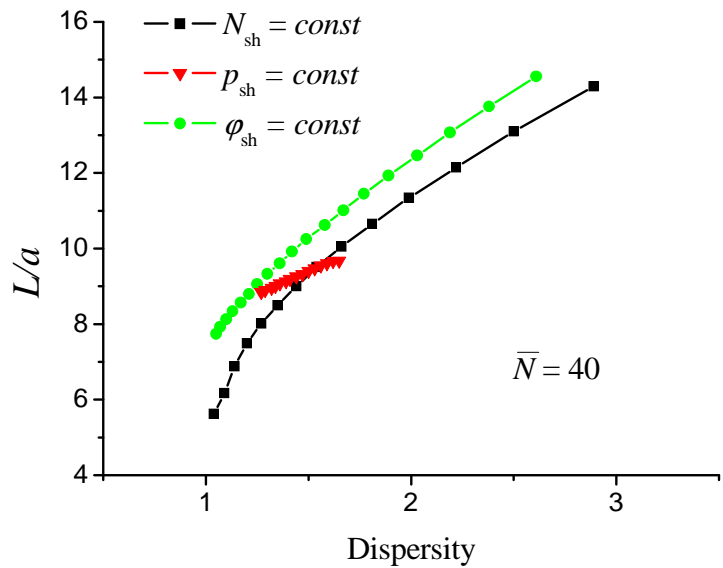

FIG. 7. Layer thickness, $L$, for the random multiblock copolymers with the fixed mean block size, $\overline{N} = 40$, vs the dispersity in the block size, $Đ$, at the fixed size of short blocks ($N_{sh}=4$, squares), fixed number fraction of short blocks ($p_{sh}=0.4$, triangles), or fixed volume fraction of short blocks ($\varphi_{sh}=0.2$, circles). The Flory-Huggins parameter, $\chi = 0.4$.

As long as the short blocks are present in the "alien" layers, the layer thickness for a random multiblock copolymer is larger than that for a regular multiblock copolymer with the block size $N=N_{long}$ (Figure 8). At larger $\chi$, for which all short blocks are pushed out, the layer thickness is less than that for a regular copolymer. To explain this difference, let us analyze the dependence of the free energy (4) on the layer thickness, $L$, at $\varphi_0=0$ and at a fixed $\chi$. The number of long blocks and their elastic contribution is only slightly less than that of a regular multiblock copolymer (by the factor $p_{long}N_{long}/\overline{N}$). The volume of the interfacial layer is approximately equal to the volume of short blocks, the interface thickness being proportional to the layer thickness (from Eqs. (1) and (2)), $D_\Delta = D+\Delta = D_0 + L\varphi_{sh}$. Then, the interaction contribution into the interfacial energy, $F_s$, is constant and the main factor is decreasing the block elastic contribution to the surface tension (the first term in Eq. (9)) and, correspondingly, the surface free energy due to a much larger interfacial layer thickness, $D_\Delta$. Note that the regular multiblock copolymer with the block size equal to that of the random one ($N=\overline{N}$ rather than $N=N_{long}$) would have a much less domain size (the dashed black curve), because it is controlled by the elasticity of long blocks rather than of blocks of size $\overline{N}$ ($\overline{N} < N_{long}$).

At very large $\chi$ and $N_{long}$, the asymptotic dependences for the regular and random multiblock copolymers can be found neglecting the entropy effects of block ends and short blocks. In this limit, the block elastic free energy can be taken in the form $f_{bl} \approx f_{gs} = \dfrac{3k_BT}{2}\dfrac{L^2}{N_{long}a^2}$. For a regular multiblock copolymer, the free energy (11) is approximately equal to $F_r \approx Mf_{gs} + \sigma_0 S$, and its minimum corresponds to $L_{a0} = (\chi/54)^{1/6} N_{long}^{2/3} a$ (more exactly, $L_{a0} - L_{str} = (\chi/54)^{1/6} N_{long}^{2/3} a$, if $f_{bl}$ is given by Eq. (7)). For a random multiblock copolymer, all "alien" short blocks are pushed out of the layers, so that $\varphi_0 = 0$ and the free energy (4) can be written as $F \approx Mp_{long}f_{gs} + \sigma S$, the interfacial tension (9) is equal to $\sigma = 0.5\sigma_0(D_0/D_\Delta + D_\Delta/D_0)$, where the interface thickness, $D_\Delta = D_0 + L\varphi_{sh}$. At $L\to\infty$, the interface free energy depends on $L$ as $F_s = \sigma S = \text{const} + 0.5\sigma_0 V(1/L + O(D_0/L^2))$. Since $V = M\overline{N}v$, then the equilibrium thickness, $L_a = L_{a0}/(2(1-\varphi_{sh}))^{1/3}$, is characterized by the same slope ($\sim \chi^{1/6}$) in the double logarithmic scale (Figure 8).

The effect of the block end entropy decreases the equilibrium layer thickness for a regular block copolymer because the additional entropy penalty $\sim \ln(L/D_0)$ per block hinders the increase in *L* (compare the blue curve with diamonds and the dashed black curve). This effect becomes negligible with the increase of $\chi$ and *N* (the decrease of the block end number).

The slope of the dependence of the layer thickness, *L*, on $\chi$ for the random multiblock copolymer is less than that for the regular copolymer (at $\chi>0.8$). This conclusion agrees with the computer simulation results,[17,18] where the very weak dependences of the structure period on the Flory-Huggins parameter were found. The dependences of *L* on $\chi$ tend to the asymptotic law $\sim\chi^{1/6}$ at $\chi N_{long}>1000$ only. Considering the dependences of the layer thickness, *L*, on the long block size one can predict that they are close to the asymptotical one $\sim N^{2/3}$ for a regular multiblock copolymer from $\chi N \approx 40$ and for a random multiblock copolymers from $\chi N_{long} \approx 100$.

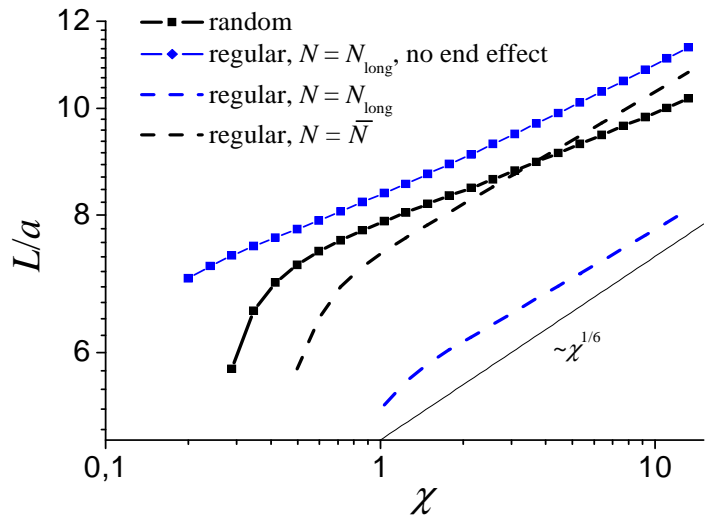

FIG. 8. Layer thickness, *L*, vs the Flory-Huggins parameter, $\chi$, in the double logarithmic scale for the random multiblock copolymer (black curve with squares, $N_{long}=40$, $N_{sh}=4$, $p_{sh}=0.4$) and for the regular block copolymers with $N=N_{long}$ (dashed thick curve) and with $N=\bar{N}=p_{long}N_{long}+p_{sh}N_{sh}=25.6$ (dashed thin curve). The blue curve with diamonds represents the layer thickness for the regular block copolymer with $N=N_{long}$ calculated neglecting the block end entropy effect. The thin line gives the asymptotic trend $\sim\chi^{1/6}$.

## IV. CONCLUSIONS

In the present work, microphase separation in random multiblock copolymers is studied using the mean-filed theory for the special type of bidisperse random multiblock copolymers with blocks of only two sizes (long and short). Such copolymer type is a model for random multiblock copolymers with high dispersity in the block size. By varying the long and short block sizes and their fractions, the mean block size and block size dispersity can be controlled.

The strong segregation regime for long blocks is assumed, whereas solitary short blocks are able to penetrate into "alien" domains and exchange between the domains and interfacial layer, where the other short blocks are concentrated. In comparison with copolymers characterized by a low polydispersity, the main features of the present consideration relate to the presence of short blocks. First, the penetration of short blocks into the "alien" domains leads to the formation of joint long blocks in their own domains. A very low elastic free energy of joint blocks, or effective enlargement of a long block size, promotes the increase of the layer thickness. Second, many short blocks are localized in the interfacial layer, which becomes much thicker in comparison with a similar melt structure for multiblock copolymers with only long blocks. Correspondingly, the elastic deformation of long blocks at the interface decreases but the interfacial interaction energy increases.

The possibility of short block penetration into the "alien" domains is controlled by the product $\chi N_{sh}$, and the volume fraction of "alien" short blocks does not exceed several per cent in terms of the present model. As a result, at not very large values of $\chi N_{sh}$ the domain size for random multiblock copolymers is larger than that for a regular copolymer consisting of the same long blocks, while at quite large values of $\chi N_{sh}$ it is smaller. The domain size increases with the polydispersity index at a fixed mean block length. The calculations were performed for a lamellar melt structure, however, the same general features can be expected for other types of microphase separation including bicontinuous-like structures.

In experiments, the structure of random multiblock copolymer melts is not strictly periodic, being characterized by only one or two peaks in their SAXS profiles. Therefore, those materials are not good, for example, for lithography. At the same time, a melt structure is often bicontinuous that is important for conducting membranes and can provide high mechanical stability. Besides, the penetration of small amount of short blocks into "alien" domains could provide the possibility to obtain sustainable plastics with both phases to be degradable. The recent developments of new experimental techniques for a synthesis of random multiblock copolymers promise new horizons in their practical applications.

## APPENDIX A: PARAMETERS OF THE INTERFACE

The interface free energy for a block copolymer melt (with long blocks only) in a strong segregation regime can be written as[33]

$$F_{s0} = (\sigma_0 + \sigma_{e0})S, \tag{A1}$$

$$\frac{\sigma_0 \upsilon}{k_B T} = a\sqrt{\frac{\chi}{6}}, \tag{A2}$$

$\sigma_0$ is the interfacial tension in a homopolymer melt, $S$ is the interface area,

$$\frac{\sigma_{e0}}{k_B T} = \frac{M_0}{S}\ln\left(\frac{M_0 \upsilon}{S e \pi D_0}\right) = \frac{M_0}{S}\ln\left(\frac{\bar{\varphi}}{e}\frac{L}{\pi D_0}\right) \tag{A3}$$

is the entropic contribution of the localization of block junction points in an interfacial region of thickness $D_0 = a/\sqrt{6\chi}$, $M_0$ is the number of block junctions, $\bar{\varphi}$ is the mean volume fraction of block junctions over the whole system (the junction point volume is equal to $\upsilon$).

The well-known expression for $\sigma_0$ (Eq. (A2)) can be obtained, for example, using the following half-empirical approach. Let $\varphi_A(x)$ be the dependence of the volume fraction of A monomer units on the coordinate $x$ along an axis perpendicular to the interface. The interfacial tension $\sigma_0 = \sigma_{el0} + \sigma_{int0}$ is equal to a sum of the elastic energy contribution, $\sigma_{el\,0}$, and contribution of the interaction energy of monomer units, $\sigma_{int\,0}$. The first term is proportional to the Lifshitz conformational entropy[42] and can be written in the form

$$\sigma_{el\,0}\upsilon/(k_B T) = \frac{a^2}{6}\int dx\left(\frac{(\varphi'_A(x))^2}{4\varphi_A(x)} + \frac{(\varphi'_B(x))^2}{4\varphi_B(x)}\right), \quad \varphi_B(x)=1-\varphi_A(x).$$ The second term is equal to $\sigma_{int\,0}\upsilon/(k_B T) = \chi\int dx\,\varphi_A(x)\varphi_B(x)$. Since the volume fractions $\varphi_A$ and $\varphi_B$ change from 0 to 1 in an interfacial region of thickness $D_0$, the values of the interfacial tension contributions are approximately equal to

$$\sigma_{el\,0}\upsilon/(k_B T) \approx \frac{a^2}{6}\frac{1}{2D_0}, \tag{A4}$$

$$\sigma_{int\,0}\upsilon/(k_B T) \approx \frac{\chi D_0}{2}. \tag{A5}$$

Minimum of the sum of the contributions (A4) and (A5) with respect to $D_0$ corresponds to the expression (A2) for the interfacial tension $\sigma_0$ at $D_0 = a/\sqrt{6\chi}$.

Now let us take into account that monomer units of "alien" type are present in all layers. The expression (A2) can be generalized using a semi-empirical approach described above. Let the volume fraction of a non-selective solvent be equal to $\varphi_0$ in all layers, the Flory-Huggins parameter of interactions between A or B monomer units and the solvent be equal to $\chi$ in all layers, and thickness of an interfacial region be equal to $D$. The total interaction energy of such system is equal to that of a random multiblock AB copolymer melt with a lamellar structure, where the volume fraction of "alien" blocks in the layers is equal to $\varphi_0$.

The interfacial tension, $\sigma_1$, without a contribution of the block end localization is equal to a sum of the elastic energy contribution, $\sigma_{el\,1}$, and contribution of the interaction energy, $\sigma_{int\,1}$: $\sigma_1=\sigma_{el\,1}+\sigma_{int\,1}$. The volume fraction of long A or B blocks changes from the value of $1-\varphi_0$ in a layer of their own type to 0 in an "alien" layer. Since the elastic free energy contribution to the surface tension is linear with respect to the block concentration at a fixed interface thickness, this contribution can be written as

$$\sigma_{el\,1}\upsilon/(k_B T) \approx \frac{a^2}{6}\frac{1-\varphi_0}{2D}. \tag{A6}$$

The interaction energy contribution to the surface tension can be estimated as

$$\sigma_{\text{int 1}} \upsilon/(k_B T) = \chi \int dx (\varphi_A(x)(1-\varphi_A(x)) - \varphi_0(1-\varphi_0)) \approx \frac{\chi(1-2\varphi_0)^2 D}{2}. \quad (A7)$$

Minimum of the sum of the expressions (A6) and (A7) for $\sigma_{\text{el 1}}$ and $\sigma_{\text{int 1}}$, respectively, over $D$ corresponds to the interface parameters

$$D = D_0 (1-\varphi_0)^{1/2} / (1-2\varphi_0), \quad \sigma_1 = \sigma_0 (1-\varphi_0)^{1/2} (1-2\varphi_0). \quad (A8)$$

Thus, the interfacial region thickness increases and the interfacial tension decreases with the growth of $\varphi_0$, $D \approx D_0(1+1.5\varphi_0)$, $\sigma_1 \approx \sigma_0(1-2.5\varphi_0)$ at $\varphi_0 \ll 1$. The approximation (A8) agrees qualitatively with the results of the self-consistent field analysis of the interfacial characteristics for a homopolymer mixture in the presence of solvent,[43,44] where the change in the solvent concentration in the interfacial region is taken into account.

In a random multiblock copolymer melt, localization of short blocks at the interface should enlarge an interfacial layer region. Let the interfacial region thickness become equal to a sum of the thickness $D$ (A8) and thickness $\varDelta$ (Eq. (1)): $D_\varDelta = \varDelta + D$. Remind that $\varDelta$ is equal to the thickness of a hypothetical layer containing only short blocks localized at the interface. Assuming that the elastic free energy contribution of long blocks and interaction energy contribution to the interfacial tension depend on the interfacial layer thickness as described by the expressions (A6) and (A7), respectively, one can calculate the interfacial tension, $\sigma$, as

$$\frac{\sigma \upsilon}{k_B T} \approx \frac{a^2}{6} \frac{1-\varphi_0}{2D_\varDelta} + \frac{\chi(1-2\varphi_0)^2 D_\varDelta}{2} \quad (A9)$$

Junction points between long blocks and a part of short blocks are localized in the interfacial region. The contribution of their localization to the interfacial tension ($\sigma_{\text{e,long}}$ and $\sigma_{\text{e,sh}}$, respectively) can be written similarly to the expression (A3):

$$\sigma_e = \sigma_{e,\text{long}} + \sigma_{e,\text{sh}},$$

$$\frac{\sigma_{e,\text{long}}}{k_B T} = \frac{M\tilde{p}_{\text{long}}}{S} \ln\left(\frac{\tilde{p}_{\text{long}}}{\overline{N}e} \frac{L}{\pi D_\varDelta}\right), \quad \tilde{p}_{\text{long}} = (1-p_b) p_{\text{long}}^2 \sum_{i=0}^{\infty} (p_{\text{long}} p_b)^i = \frac{(1-p_b) p_{\text{long}}^2}{1-p_{\text{long}} p_b}$$

$$\frac{\sigma_{e,sh} S \upsilon}{k_B T V} = \frac{\varphi_1 - \varphi_0}{N_{sh}} \ln\left(\frac{\varphi_1 - \varphi_0}{e} \frac{L}{\pi D_\Delta}\right) + \frac{\varphi_{sh}}{N_{sh}} p_{long}^2 \sum_{i=2}^{\infty} p_{sh}^{i-1} \ln\left(\frac{\varphi_i}{e} \frac{L}{\pi D_\Delta}\right),$$

where $M\tilde{p}_{long}$ is the total number of block ends for long joint blocks beginning at the interface and ending at the junction with a long block (two long blocks of the same type separated by a short block in an "alien" layer belong to one joint block), the relation $V=SL=M\overline{N}\upsilon$ being taken into account. Short blocks adjacent in the chain are assumed to be able to move as one object. The first term in the short block contribution describes solitary short blocks (adjacent to two long blocks), $\varphi_i = i p_{sh}^{i-1} p_{long}^2 \varphi_{sh}$.

## APPENDIX B: ELASTIC FREE ENERGY OF BLOCKS

Let us consider a polymer block consisting of $k$ monomer units in a layer of thickness $L$, both ends of the block being localized at an interface. Let $L_0 = \sqrt{k}a$ be the characteristic spatial size of the block with a completely random conformation. If the layer thickness is not large ($L<L_0$), then blocks are not stretched and their elastic free energy can be calculated as an additional contribution to the free energy of a homogeneous melt, as was done for the elastic free energy of blocks in a polymer globule with both ends at its surface.[41] The elastic free energy of a block calculated using this approach is denoted by $f_L(k,L)$.

Taking into account that both "loop" and "bridge" conformations are possible, the block elastic free energy can be written as

$$\frac{f_L(k,L)}{k_B T} = -\ln(G_1(k,L)a + G_2(k,L)a), \tag{B1}$$

where $G_1(k,L)$ and $G_2(k,L)$ are the statistical weights of a "loops" and "bridge", respectively. The sum $G_1+G_2$ is proportional to the number of conformations of a block with the ends at the layer surfaces.

To find $G_1(k,L)$ and $G_2(k,L)$, it is necessary to calculate the Green function of a block, $G(\mathbf{r},k|\mathbf{r}_0)$, consisting of $k$ statistical segments of length $a$ with the beginning at the point

$\mathbf{r_0} = (x_0, y_0, z_0)$ and end at the point $\mathbf{r} = (x,y,z)$ of the layer. The Green function describes a random block conformation in the layer and satisfies the diffusion-type equation[42,45]

$$\frac{\partial}{\partial k} G(\mathbf{r}, k | \mathbf{r_0}) = \frac{a^2}{6} \Delta G(\mathbf{r}, k | \mathbf{r_0}), \quad \mathbf{r}, \mathbf{r_0} \in V_L, \tag{B2}$$

where $\Delta$ is a Laplacian with respect to $\mathbf{r}$, $V_L$ is the layer of thickness $L$. The initial condition is

$$G(\mathbf{r}, 0 | \mathbf{r_0}) = \delta(\mathbf{r} - \mathbf{r_0}). \tag{B3}$$

The boundary conditions are

$$\left. \frac{\partial G}{\partial x} \right|_{x=0} = \left. \frac{\partial G}{\partial x} \right|_{x=L} = 0. \tag{B4}$$

A zero probability flux at the planes $x=0$ и $x=L$ (B4) corresponds to the condition of a constant polymer density in the layer.

The solution of the equation (B2) under the conditions (B3) and (B4) can be written as a product

$$G(\mathbf{r}, k | \mathbf{r_0}) = G(x, k | x_0) G_0(y, k | y_0) G_0(z, k | z_0), \tag{B5}$$

where the Green function $G(x, k | x_0)$ is equal to

$$G(x, k | x_0) = \frac{1}{L} + \frac{2}{L} \sum_{m=1}^{\infty} \cos \frac{\pi m x_0}{L} \cos \frac{\pi m x}{L} \exp\left(-\frac{a^2 \pi^2 m^2 k}{6 L^2}\right). \tag{B6}$$

The Green functions $G_0(y, k | y_0)$ and $G_0(z, k | z_0)$ describe a free random walk along the axes $y$ and $z$, they are given by the expression $G_0(y, k | y_0) = \sqrt{\frac{3}{2\pi k a^2}} \exp\left(-\frac{3}{2 k a^2} (y - y_0)^2\right)$.

For a "loop" conformation, the beginning and end coordinates are $x = x_0 = 0$, for a "bridge" conformation $x_0 = 0$, $x = L$. At $k >> (L/a)^2$, the first term dominates in the sum (B6) and the statistical weights of "loops" and "bridges" can be estimated as $G_1(k, L) \approx \frac{1}{L}\left(1 + 2\exp(-a^2 \pi^2 k / 6 L^2)\right)$ and $G_2(k, L) \approx \frac{1}{L}\left(1 - 2\exp(-a^2 \pi^2 k / 6 L^2)\right)$, respectively. At $1 << k << (L/a)^2$, the sum can be replaced by an integral and then $G_1(k,L) \approx \frac{1}{L} + \sqrt{\frac{6}{\pi k a^2}} \approx \sqrt{\frac{6}{\pi k a^2}}$. More exactly, the statistical weights of "loops" and "bridges" for discrete values of block lengths are calculated as $G_1(k,L) = \int_{k-1}^{k} dk\, G(0,k\,|\,0)$ and $G_2(k,L) = \int_{k-1}^{k} dk\, G(L,k\,|\,0)$, respectively,

$$G_1(k,L) = \frac{1}{L} + \frac{12L}{a^2\pi^2}\sum_{m=1}^{\infty}\frac{1}{m^2}\left(\exp\left(-\frac{a^2\pi^2 m^2(k-1)}{6L^2}\right) - \exp\left(-\frac{a^2\pi^2 m^2 k}{6L^2}\right)\right)$$

$$G_2(k,L) = \frac{1}{L} + \frac{12L}{a^2\pi^2}\sum_{m=1}^{\infty}\frac{(-1)^m}{m^2}\left(\exp\left(-\frac{a^2\pi^2 m^2(k-1)}{6L^2}\right) - \exp\left(-\frac{a^2\pi^2 m^2 k}{6L^2}\right)\right)$$

The dependences of the block elastic free energy (Eq. (B1)), $f_L$, on the block size at the fixed layer thickness and on the layer thickness at a fixed block size are presented in Figures B1 and B2, respectively. The elastic free energies of "loops", $f_{loop}/(k_B T) = -\ln(G_1(k,L)a)$, and "bridges", $f_{br}/(k_B T) = -\ln(G_2(k,L)a)$, are shown as well. All block free energies increases with $L$, $f_{loop}$ and $f_L$ become almost constant $L > L_0$ and $L > 2L_0$, respectively. At $k < (L/a)^2$, the elastic free energy of "bridges" is considerably larger than that of "loops" and these energies become practically equal to each other for $k > 2(L/a)^2$. The fraction of "loops" (blocks that begin and end at the same layer surface) $p_1$ and "bridges" (blocks that begin and end at the different surfaces of the layer) $p_2$ can be directly calculated from the Green functions $G_1(k,L)$ and $G_2(k,L)$, which are proportional to the number of corresponding conformations, $p_1 = \frac{G_1(k,L)}{G_1(k,L) + G_2(k,L)}$, $p_2 = \frac{G_2(k,L)}{G_1(k,L) + G_2(k,L)}$.

The "bridge" fraction $p_2$ decreses from approximately 0.5 at $L \sim a$ to 0.3 at $L = L_0$.

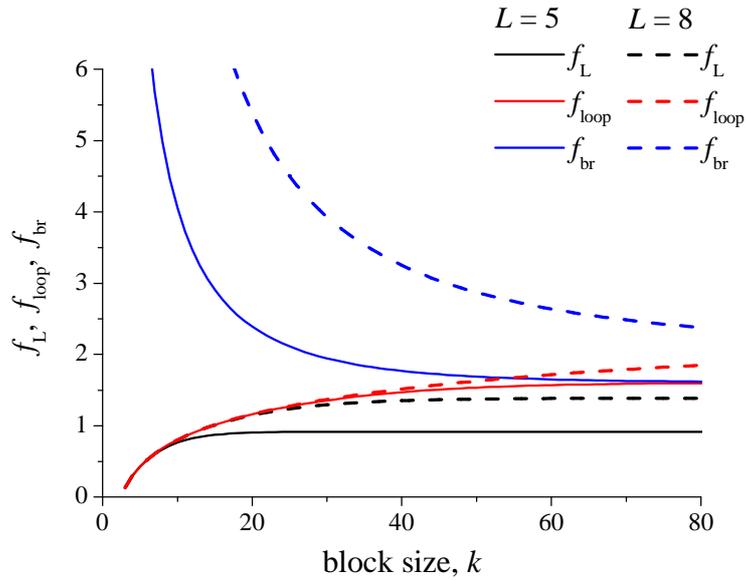

FIG. 9. Elastic free energy of a block, $f_L$, in units $k_BT$, vs the block size, $k$, at the layer thickness $L=5a$ and $8a$ (the black curves). The elastic free energies of "loop" and "bridge" blocks, $f_{loop}$, and $f_{br}$, are shown by the red and blue curves, respectively. Solid curves are plotted for $L=5a$ and dashed ones for $L=8a$.

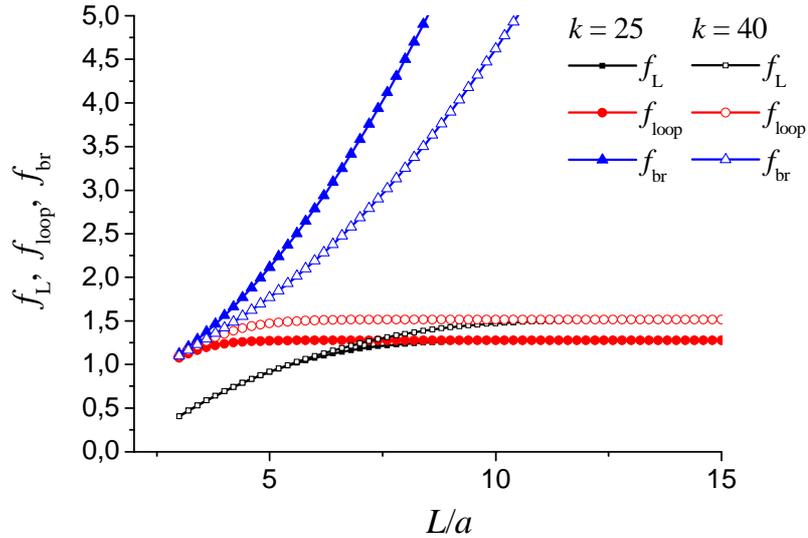

FIG. 10. Total elastic free energy of a block, $f_L$, in units $k_BT$, vs the layer thickness, $L$, for blocks of size $k=25$ and 40 (the black curves). The elastic free energies of "loop" and "bridge" blocks, $f_{loop}$, and $f_{br}$, are shown by the red and blue curves, respectively. Solid symbols are used for $k=25$ and open ones for $k=40$.

## ACKNOWLEDGMENTS

We appreciate the financial support from the Russian Science Foundation (project 14-13-00683).